# The LHC based µp colliders

Umit Kaya[a,b], Bora Ketenoglu[b], Saleh Sultansoy[a,c]

[a] *TOBB University of Science and Technology, 06500 Sogutozu, Ankara, Turkey*
[b] *Ankara University, 06100 Tandogan, Ankara, Turkey*
[c] *ANAS Institute of Physics, Baku, Azerbaijan*

## Abstract

Construction of muon collider or dedicated muon ring tangential to the LHC will give opportunity to handle µp collisions at multi-TeV center of mass energies. Main parameters and physics search potential of the LHC based µp colliders are discussed.

## 1. Introduction

TeV energy muon-proton colliders [1, 2] were proposed twenty years ago as alternative to Linac-HERA and Linac-LHC based ep and γp colliders (see review [3] and references therein). Two years later an ultimate √s=100 TeV µp collider (with additional 50 TeV proton ring in √s = 100 TeV muon collider tunnel) was suggested in [4]. It should be noted that luminosities of µp collisions in [1] and [2] were over-estimated (see sub-sections 3.2 and 2.3 in [4], respectively). Recently, FCC and SppC based energy frontier muon-hadron colliders have been proposed in [5] and [6], respectively.

In this paper we consider main parameters and physics search potential of the LHC based µp colliders. Center of mass energies and luminosities for different options are estimated in Section 2. Then, in Section 3 we briefly discuss physics at these colliders. Finally, in Section 4 we present our conclusions and recommendations.

## 2. Luminosity estimations

For the LHC proton beam parameters we use High Luminosity upgraded values from [7] presented in Table 1.

Table 1: Main parameters of proton beam

| Parameter | HL-LHC |
|---|---|
| proton energy [TeV] | 7 |
| bunch intensity | $2.2 \times 10^{11}$ |
| normalized rms emittance [mm] | 2.0 |
| regular bunch spacing [ns] | 25 |
| no. bunches per ring | 2760 |
| β* possible for e-p IP [m] | 0.07 |
| Beam size at IP [µm] | 4.45 |

Muon collider's parameters taken from recent paper [8] are presented in Table 2.

Table 2: Table 5: Muon collider design parameters

| Parameter | | | |
|---|---|---|---|
| Muon energy, TeV | 0.75 | 1.5 | 3 |
| Repetition rate, Hz | 15 | 12 | 6 |
| Average luminosity per IP, $10^{34}$cm$^{-2}$s$^{-1}$ | 1.25 | 4.6 | 11 |
| Circumference, km | 2.5 | 4.34 | 6 |
| β*, cm | 1 | 0.5 | 0.3 |
| Normalized emittance, π·mm·mrad | 25 | 25 | 25 |
| Bunch length, cm | 1 | 0.5 | 0.3 |
| Number of muons per bunch, $10^{12}$ | 2 | 2 | 2 |
| Number of bunches per beam | 1 | 1 | 1 |
| Beam-beam parameter per IP | 0.09 | 0.09 | 0.09 |
| Beam size at IP, μm | 6 | 3 | 1.64 |

Using corresponding formulas from [5] and [6] one can obtain center of mass energy and luminosity values for the LHC based μp colliders presented in Table 3.

Table 3: Main parameters of the LHC based μp colliders

| Parameter | | | |
|---|---|---|---|
| Muon energy, TeV | 0.75 | 1.5 | 3 |
| √s, TeV | 4.58 | 6.48 | 9.16 |
| Luminosity, $10^{33}$cm$^{-2}$s$^{-1}$ | 1.4 | 2.3 | 0.9 |

## 3. Physics search potential

Because of high center of mass energy and luminosity values the LHC based μp colliders have a huge potential for the SM and BSM searches. Conserning SM physics, they will provide precision PDF's for the FCC and SppC (100 TeV center of mass energy pp colliders planned for 2030'ies). Small $x_g$ region, which is crucial for understanding of QCD basics, can be explored down to $10^{-8}$. Precision Higgs physics is another important topic, which should be analysed in details…

Concerning BSM physics, LHC based μp colliders are comparable or essentially exceeds potential of the LHC itself in a lot of topics, such as leptoquarks related to the second family leptons, excited muon, excited muon neutrino, colour octet muon, contact interactions, SUSY, RPV SUSY (especially resonant production of corresponding squarks), extended gauge simmetry etc.

## 4. Conclusions

It is shown that construction of future muon collider (or dedicated μ-ring) tangential to the LHC will essentially enlarge physics search potential for both the SM and BSM phenomena. Therefore, systematic study of accelerator, detector and physics search aspects of the LHC based µp colliders is necessary for long-range planning of HEP.

In principle one can consider two-stage scenario for the LHC based lepton-hadron colliders: the LHeC option with 9 km e-ring [9] as the first stage, following by construction of µ-ring in the same tunnel.